\documentclass[pra]{revtex4}

\begin{document}
\title{All tight multipartite Bell correlation inequalities for three dichotomic observables per observer}
\author{M. \.Zukowski}
\affiliation{Institute of Theoretical Physics and Astrophysics, Uniwersytet
Gda\'nski, PL-80-952 Gda\'nsk}

\maketitle

{\small  Abstract: {A derivation 
of the full set of Bell inequalities involving correlation functions, for two parties, with  binary observables, and three
possible local settings. The procedure can be extended straightforwardly to multiparty correlations.}}
\section{Introduction}

This paper contains a study of  some technical aspects associated with multiparty Bell
inequalities. Multiparty Bell inequalities are recently gaining in importance, because security, or performance, of many of quantum communication schemes
(like multiparty key  (secret) sharing \cite{SCARANIGISIN}, quantum communication complexity problems \cite{BZPZ}) can be measured with Bell inequalities of some form \cite{USEFUL_ENT}. Bell inequalities show the limit of achievable correlations between many parties, if they share only classical means of communication and/or 
computation. If violated by certain predictions for a certain  entangled state, they show that this state, without any further manipulations is a good resource for some quantum informational tasks. To put is short, in the standard context Bell inequalities show that if there are $N$ separated parties, each equipped with a classical supercomputer, who share common data, and perhaps programs, but otherwise are not allowed to communicate, the locally run programs cannot 
simulate some phenomena associated with spatially separated $N$ local measurements, with locally decided settings, on $N$ entangled particles, when each party is asked to predict (compute) result of a measurement for a different particle.

Bell's theorem was formulated for two particles \cite{BELL64}. It took a quarter of century to realize that for more than two particles the situation is much more interesting. The discovery of Greenberger-Horne-Zeilinger correlations \cite{GHZ} immediately led to various generalizations of the multipartite problem. 
This gave birth to the quest of finding Bell inequalities for the new problems. Mermin was first to produce series of inequalities for arbitrary many particles, involving dichotomic observables, and allowing each observer to choose between two settings \cite{MERMIN}.
A complementary series of inequalities was introduced by Ardehali \cite{ARDEHALI}. In the next step Belinskii and Klyshko gave series of two settings inequalities, which contained the tight inequalities of Mermin and Ardehali \cite{BELINSKII-KLYSHKO}. Finally the full set of tight two setting per observer, $N$ party Bell inequalities for dichotomic observables was found independently in \cite{WW, WZ,ZB}. All these series of inequalities
are a generalization of the CHSH ones \cite{CHSH}. Such inequalities involve only $N$ party correlation functions. The aim of this paper is to go one step further, and to produce series of tight CHSH-type Bell inequalities for $N$ observers, dichotomic observables, and {\em{three}} possible setting for each observer. Recently series of tight inequalities, which do not form a complete set, were found for the case when $k$-th partner can  choose between $2^{k-1}$ settings (see \cite{WUZONG1, WUZONG2, LPZB}). We shall not follow this approach here, as it seems to be incapable of generating
the full set of Bell inequalities. Also the methods of \cite{KWEK} and \cite{ROT} will be not discussed here, as they do not follow from the analysis of the geometry of the polytope of local realistic models (for this concept see \cite{PITOVSKY}). 

Only the proof for the necessary and sufficient condition for the two-observer ($3\times3$) case will be shown in full detail. As the reader will see, the generalization 
to more parties, that is e.g. to $3\times3\cdots\times3$ problems, is straightforward. In all these cases we shall be concerned with the generalization of the inequalities of the CHSH type. Such inequalities involve only $N$ party correlation functions.
However, a remark will be made on how to extend the validity of the presented results to CH-type inequalities \cite{CH} (which involve also correlations of lower rank). We shall study only experiments involving dichotomic observables.

We have chosen the three settings case, with two valued observables, because
it is the simplest example for which the method used here can lead to new results.
Also the specific trait of three settings inequalities is that the observers, in the case of qubits, can use a full set of mutually unbiased 
observables for their measurements. The aim of this paper is basically a presentation of full details of the derivation of the form of the coefficients in three setting inequalities. Explicit forms of inequalities that can be obtained using this formalism will be presented elsewhere. 

Finally note that  the results of Garg \cite{GARG} are equivalent to the present ones, for the two observer case. However, the method 
presented here is easily generalizable to the case of three or more particles, and therefore extends to an unknown territory. Some results specific results concerning three-particle inequalities are presented in  \cite{WIESNIAK-ZUK}. 

\section{Definition of  $3\times3$ problem}

We shall be interested only in the conditions of existence of local realistic models of the 
two particle correlation functions, forgetting totally the lower order (local) processes.
A proper symmetrization of a solution of the present problem, leads always to a solution of 
a problem in which one assumes that lower order (one particle) averages vanish.
We allow each 
observer to choose between three settings. We will search for the positivity 
condition for the unphysical (hidden) probability distribution involving results for all settings:
\begin{equation}
P(m_1^{(0)},m_1^{(1)},m_1^{(2)},m_2^{(0)},m_2^{(1)},m_2^{(2)}),
\end{equation}  
where $m_i^{(n_i)}$ is the local realistic prediction for measurement result
if the observer $i$ chooses to measure a dichotomic observable number $n_i$.
This distribution must properly reproduce all values of two particle correlation function for the full set of 
$3\times3$ experiments.

\subsection{Structure of the polytope}
The classical local realistic model of the two particle correlations is 
essentially a probabilistic mixture of deterministic "orders" that are sent by the source to the 
detection regions. The orders sent to one of the observation stations, in the case of three possible settings,
must contain the information which value of the local dichotomic observable is to
result during the measurement. Therefore such an order, for the first observer, 
can be put in the following way:
\begin{equation}
(m_1^{0}, m_1^{1}, m_1^{2})=(\pm1,\pm1,\pm1).
\end{equation}
We have $2^3$ distinct orders, that can be sent to one of the observation stations. 

If one is interested only in local realistic models of the 
two-particle (highest order) correlation functions, then the optimal way to describe 
the deterministic orders is by tensor products of the orders sent to each of the observers.
Thus the deterministic orders are now:
\begin{equation}
(m_1^{0}, m_1^{1}, m_1^{2})\otimes(m_2^{0}, m_2^{1}, m_2^{2})=(\pm1,\pm1,\pm1)\otimes(\pm1,\pm1,\pm1).
\label{ORDER}
\end{equation}
Note however that, the description has an in-built simplification, because one cannot distinguish between  
\begin{equation}
(m_1^{0}, m_1^{1}, m_1^{2})\otimes(-m_2^{0}, -m_2^{1}, -m_2^{2})
\end{equation}
and
\begin{equation}
(-m_1^{0}, -m_1^{1}, -m_1^{2})\otimes(m_2^{0}, m_2^{1}, m_2^{2}).
\end{equation}

A local and realistic model of the correlation functions $E_{n_1n_2}$, $n_1, with n_2=0,1,2$, exist if the 
``tensor" built out of these values of the correlation function, $\hat{E}$, is reproduced by 
the model of the following structure:

\begin{eqnarray}
&\hat{E}=\sum_{m_1^{(0)},m_1^{(1)},m_1^{(2)},m_2^{(0)},m_2^{(1)},m_2^{(2)}=\pm1}P(m_1^{(0)},
m_1^{(1)},m_1^{(2)},m_2^{(0)},m_2^{(1)},m_2^{(2)})&\nonumber\\
&
\times(m_1^{0}, m_1^{1}, m_1^{2})\otimes(m_2^{0}, m_2^{1}, m_2^{2}).&\nonumber\\
\label{MODEL}
\end{eqnarray}

As we shall see, the tensor product structure of the full set of correlation functions is a very important property, which can be used to simplify the derivations. This one of the basic concepts, around which this paper is formulated. 

\section{Pedagogical interlude: structure of polytope of models for just one side}

In this subsection are presented basic the traits of realistic models 
for results of experiments on just one side (say for Alice). The full set of such models, as we shall see, forms a three dimensional cube. The aim of this section is to get some intuition, before 
one moves to the much more complicated case of two separated observers, and the problem of the local realistic models for their observations. An impatient reader can move right away to the next section.

If one is interested only in the events on one side, say Alice's, of the experiment,
then all possible realistic models of the average values obtained in the three 
possible experiments (settings) are convex (i.e., probabilistic) combinations
of the local orders (\ref{ORDER}). Therefore any realistic model lays within the interior of 
a cube, with eight vertices defined by (\ref{ORDER}).

Now let us understand some of the traits of this cube. First notice that the six faces contain
the following vertices:
the first face 
\begin{equation}
(1,a,b),
\end{equation}
with $a,b=\pm1$,
the second face, on the opposite side
\begin{equation}
(-1,a,b),
\end{equation}
the third one\begin{equation}
(a,1,b),
\end{equation}
the fourth
\begin{equation}
(a,-1,b),
\end{equation}
fifth
\begin{equation}
(a,b,1),
\end{equation}
and the last one
\begin{equation}
(a,b,-1).
\end{equation}
Always three arbitrary vertices of one face form a complete set in $\bf R^3$ (a non-orthonormal basis).
The missing fourth vertex from the given face, once one chooses the three vertex basis, is a linear combination of the basis
vertices, {\em with the expansion coefficients of modulus of one, and the sum of the coefficients equal to one too}.
For example take the first face. The four vertices satisfy the following simple relation:
\begin{equation}
\sum_{a,b=\pm1}ab(1,a,b)=0.
\label{LINEAR-DEPENDENCE}
\end{equation} 
But one does not have to manipulate this relation in order to make sure that the italicized thesis is correct.
Simply there is a well known fact from analytic geometry that makes this relation obvious:
{\em A hyperplane in an N dimensional real space is defined by N linearly independent points that belong to it, and all other points belonging to this
hyperplane are linear combinations of these N points, with coefficients that have the property that their sum is equal to one}.
That is if $\vec{r}_i$, with $i=1,...,N$, belong to a hyperplane and are linearly independent, then any other point of the hyperplane $\vec{r}$ is expressible by
$\vec{r}=\sum_i t_i\vec{r}_i$, with  $\sum_i t_i=1$.

Just to build some intuitive background for the more complicated case to be considered in the sequel, please notice the following
simple facts about the cube. 

If we take at random three vertices, then these may belong to one face. If this is so, there is a fourth one which belongs to this face. Consequently,
the plane defined by the three vertices contains a face of the (convex) polytope. There are six such planes, and therefore 
six inequalities (defining half spaces), that have the property that the points satisfying all these inequalities form our cube (these could be called local 
Bell inequalities). 

If the three randomly chosen vertices do not belong to a single face, then they define a plane that cuts through the cube.
This in turn means that there is a pair of vertices $\vec{v}$ and $-\vec{v}$, which has the property that the vertices are in the interiors 
of the two half spaces defined by the plane, each one in a different one. 

Next, a trivial but important observation. A random triplet
of vertices does not have to form a complete set. In such a case, there is a vertex $\vec{v}$ inexpressible by a linear combination of the trio, 
and the vertex $-\vec{v}$ has the same property.

Finally, it is easy to see that the relation (\ref{LINEAR-DEPENDENCE}) has the following corollary. If one builds a (nonorthonormalized) basis out of three vertices of the cube, then all vertices of the cube are linear combinations of the basis vertices with coefficients equal to $\pm1$, or zero.
As a matter of fact either three of the coefficients have modulus equal to one, or just one of them.

As we shall see below, the polytope of local realistic models for two-observer experiments has many traits which are generalizations of the ones presented above for the cube. These traits can be used to 
determine the full set of (tight) Bell inequalities satisfied by the local realistic models. 

\section{Back to the correlation polytope}

In this section the general properties of the faces of the correlation polytope will be shown. These properties define 
the hyperplanes which contain the faces of the polytope. Therefore they define the Bell inequalities (the full set of tight Bell inequalities).

Now we work in a 9 dimensional real space of vectors. The correlation polytope has 32 vertices. Each vertex $v$ has its twin $-v$.
Thus the set of vertices can always be split into two disjoint subsets of 16, 
such that if $v$ belongs to one of them $-v$ belongs to the other one. Each such a subset contains subsets of nine vertices which form a non-orthonormalized 
basis. The question is what is the relation of the hyperplane, defined by the nine vertices, with the polytope. Does it contain a face of it, or does it cut 
through it? In the first case the equation of the hyperplane, in an obvious way, leads to a Bell inequality.

{\em Fact 1.} Let us take a random set of nine vertices, $v_1$, $v_2$,...,$v_9$, which forms a complete non-orthonormal basis.
Assume that this set has a property that 7 more vertices are linear combinations of
the nine basis ones, with the sum of all expansion coefficients equal to $+1$. Thus, because of the property of the hyperplane,
all 16 vertices are in the same hyperplane. Thesis: no other vertex is in the hyperplane, and all those other vertices are on one side of 
the hyperplane (i.e., in the interior of a single half space, out of the two which are defined by the splitting of the space by the hyperplane).

Before we start with the proof, let us have a look at a well known fact from analytic geometry:
The equation of a hyperplane in a N dimensional space, defined by N linearly independent points $w^1$, ... $w^N$, is given by following relation. 
The point $x$ belongs to the hyperplane if
\begin{equation}
D[x,w^1,w^2,...,w^N]=det\left[
\begin{array}{cccccc}
1 & x_1 & x_2 &...& x_N  \\
1 & w^1_1 & w^1_2&...& w^1_N \\
...\\
1 & w^N_1 & w^N_2 &...& w^N_N\\
\end{array}
\right]=0.\label{DETERMINANT}
\end{equation}  
$D[x,w^1,w^2,...,w^N]$ stands for the determinant of a $(N+1)\times(N+1)$ matrix with the first row beginning with $1$, and then containing 
all N coordinates of the point $x$, that is $x_1$,..., $x_N$, the $k$-th row, $k=2,3,...N$, also beginning with $1$, and then containing the coordinates of the point $w^k$ (placed in the same order as those of $x$).
Note that the first column of this matrix has all entries equal to $1$. The two half-spaces, with the hyperplane as their joint subset,  are defined by 
$D[x,w^1,w^2,...,w^N]\leq0$
and
$D[x,w^1,w^2,...,w^N]\geq0$.

Let us now move to the proof of Fact 1. The vertex, $v_x$, not belonging to the hyperplane, which contains 16 vertces, may be either equal to some $-v_k$, or is a linear combination
 $\sum_{i=1}^9t_iv_i$, with $\sum_{i=1}^9t_i=-1$. In the first case, one can,  assume first that $v_x=-v_1$. Then
 $
 D[-v_1,v_1,v_2,...,v_9]
 $
 can be expanded using the minors associated with the first column. Therefore it reads
 \begin{eqnarray}
 &d[v_1,v_2,...,v_9]-d[-v_1,v_2,...,v_9]+d[-v_1,v_1, v_3,v_4,...,v_9]& \nonumber \\
 &....-d[-v_1,v_1,v_2,...,v_8],&\nonumber\\ 
 \end{eqnarray}
where $d[v_1,v_2,v_3,..., v_9]$ is the determinant of a $9\times9$ matrix with rows defined by the vectors $v_1$, ... $v_9$. Of course, such a determinant does not vanish only if all rows of the matrix are linearly independent. Thus 
 \begin{equation}
 D[-v_1,v_1,v_2,...,v_9]=2d[v_1,v_2,...,v_9].
 \end{equation}
Similarly, if $v_x=-v_2$, then 
\begin{equation}
D[-v_2,v_1,v_2,...,v_9]=2d[v_1,v_2,....,v_9].
\end{equation}
Generally,
\begin{equation}
D[-v_k,v_1,v_2,...,v_9]=2d[v_1,v_2,....,v_9].
\end{equation}
That is, all points $-v_k$ are in the interior of only one of the two half-spaces.
Let us therefore move to the second case of $v_x=\sum_{i=1}^9t_iv_i$, with $\sum_{i=1}^9t_i=-1$.
\begin{eqnarray}
&D[\sum_{i=1}^9t_iv_i,v_1,v_2,...,v_9]=d[v_1,v_2,....,v_9]-t_1d[v_1,v_2,...,v_9]& \nonumber \\
&+t_2d[v_2,v_1,v_3,v_4,...,v_9]+...-t_9d[v_9,v_1,v_2,...,v_8].&\nonumber \\
\end{eqnarray}
But, since the opposite vertex $-v_x= -\sum_{i=1}^9t_iv_i$, by our current assumption, belongs to the hyperplane,
one has
\begin{eqnarray}
&0=D[-\sum_{i=1}^9t_iv_i,v_1,v_2,...,v_9]=d[v_1,v_2,....,v_9]-\big(-t_1d[v_1,v_2,...,v_9]& \nonumber \\
&+t_2d[v_2,v_1,v_3,v_4,...,v_9]+...-t_9d[v_9,v_1,v_2,...,v_8]\big).& \nonumber \\
\end{eqnarray}
therefore again we have
\begin{equation}
D[v_x,v_1,v_2,...,v_9]=2d[v_1,v_2,....,v_9].
\end{equation}
QED.

{\em Fact 2.} If vertices $v_r$ and $-v_r $ belong to a hyperplane, then the hyperplane contains the central point
of the polytope, and thus it cuts through the polytope.

Before we proceed further, it is good to notice the following simple property of the vertices of the polytope. 

{\em Fact 3.} Every vertex $v_q$ is a linear combination of nine vertices $v_k$, with $k=1,2,...,9$, which form a complete non-orthonormal basis (this is obvious), $v_q=\sum^9_{k=1}t_kv_k$. Take any  vertex outside of this set of nine, i.e. certain $v_x$, such that $v_x\neq\pm v_k$. Then, the expansion coefficients of $v_x$ in terms of the basis vertices have their moduli either equal to zero, or to a natural number.\footnote{As a matter of fact one can put a conjecture that $|t_k|=0$, or $1$, but proving this is not needed for the derivation of the general form of the inequalities. As such a proof seems to be cumbersome it is left as an open problem.}
Note that this is an extension of one of the properties of the cube.

Proof. Take nine vertices of the form $(1,a',b')\otimes(1,c',d')$ with $a',b',c',d'=\pm1$ ,
 but with the following cases excluded $a'=b'=1$ and $c'=d'=1$. The prime is to indicate that not 
 all values $\pm1$ are allowed.
Since $(1,1,1)=-\sum_{(a',b')\neq(1,1)}a'b'(1,a',b')$, the immediate consequence is that $(1,1,1)\times(1,c',d')$ have the  required property. So do
$(1,a',b')\otimes(1,1,1)$. Finally $(1,1,1)\times(1,1,1)=\sum_{(a',b')\neq(1,1),(c',d')\neq(1,1)}a'b'c'd'(1,a',b')\otimes(1,c',d')$ has also the required property.
All other vertices are the one listed above with an overall minus sign in
 front of them, thus for them the property holds too. Thus in such a case all expansion coefficients have the property: $|t_k|=0$ or 1.
Take an other set of  nine vertices, which form a complete set. In the simplest case, such a  new basis can also be a  ``product basis". E.g., it can be defined by, say, excluding cases like, e.g. $a"=b"=1$ and $c"=-d"=1$, and assigning to the chosen vertices specific signs, 
i.e. the basis would now be $\sigma(a",b",c",d")(1,a",b")(1,c",d")$for $(a",b")\neq(1,1)$ and $(c",d")\neq(1,-1)$, where 
$\sigma(a",b",c",d")=\pm1$ is a "sign" function. This means that with respect to the basis, $(1,a",b")\otimes(1,c",d")$, for which the required property obviously holds,
some of the expansion coefficients would change their sign, {\em but not the modulus}.  That is, they again are equal to $\pm1$ or $0$. Further, one may have a set of nine linearly independent vertices, which cannot be expressed as a set of tensor products of two three dimensional bases. E.g., this could be the first basis used in this proof with, say,
 $(1,-1,-1)\otimes(1,-1,-1)$ replaced by $(1,-1,-1)\otimes(1,1,1)$. Therefore now $(1,-1,-1)\otimes(1,-1,-1)$ is outside the basis. But it can be expressed as $(1,-1,-1)\otimes[-(1,1,1)+(1,-1,1)+(1,1,-1)]$, i.e. the expansion coefficients are of modulus $0$ or $1$. Other vectors, which did not belong to the old basis, and do not belong to the new one, have to be re-expressed in a similar way, whenever they were a linear combination which included $(1,-1,-1)\otimes(1,-1,-1)$. 
 It is therefore obvious that their new expansion coefficients can only: stay unchanged, or change their values by $\pm1$. Finally, if one considers one more replacement in the basis, this again can change the expansion coefficients by only $\pm1$. Therefore, their moduli must be  equal to $0,1,2,...$.
 QED.

{\em Fact 4.} Assume now that we choose at random nine vertices, $v_k$, with $k=1,2,...,9$, and they turn out to form a complete
non-orthonormal basis. 
No pair $v_r$ and $-v_r$ belongs to the hyperplane defined by the vertices. This is due to the fact, that if the first vertex belongs to the hyperplane, the sum of its expansion coefficients is $1$, and thus for the opposite vertex it must be $-1$, i.e. it is out of the hyperplane.  
Therefore if $v_i$ belongs to the hyperplane, then $-v_i$ is out of it. As it was shown earlier, such hyperplane can contain maximally 16 vertices. If it contains 15 or less vertices,  there must be a pair $v_x$ and $-v_x$
which does not belong to it. We shall show that in such a case the hyperplane cuts the interval defined by $-v_x$
and $v_x$ into two non-trivial intervals, and therefore it cannot be a hyperplane that contains  a face of the polytope. That is equivalent to the statement, that the hyperplane cuts through the polytope, and inequalities associated with it 
cannot be used to define the interior of the polytope (i.e., they are not Bell inequalities).

Proof. Let us calculate the value of $D[ \pm\sum_{i=1}^9t_iv_i,v_1,v_2,...,v_9]$. If one expands this determinant into determinants of minors
associated with the first column, one gets
\begin{eqnarray}
&D[\pm\sum_{i=1}^9t_iv_i,v_1,v_2,...,v_9]=d[v_1,v_2,....,v_9]\pm\big(-t_1d[v_1,v_2,...,v_9]& \nonumber \\
&+t_2d[v_2,v_1,v_3,v_4,...,v_9]+...-t_9d[v_9,v_1,v_2,...,v_8]\big)&\nonumber \\
&=d[v_1,v_2,....,v_9](1\mp\sum_it_i). \label{OUTSIDE}
\end{eqnarray}
But since it was assumed that neither $v_x=\sum_{i=1}^9t_iv_i$ nor $-v_x$ belongs to the hyperplane defined by $v_k$'s, one must have $|\sum_{i=1}^9t_i|\neq 1$. It will be  shown that  $|\sum_it_i|\geq 3$. 

First, notice that either $v_x$ or $-v_x$  belongs to the hyperplane $H_o$ containing all vertices of the 
form $(1,a,b)\otimes(1,c,d)$ (let us call this set $F_o$). Next consider a randomly chosen set of any nine {\em linearly independent} vertices, $B$. 
It can be always constructed in the following way. We choose in $F_o$ a set $B_o$, which contains nine linearly independent vertices, and then choose a random subset of these, and we multiply them by $-1$. I.e., the set $B$ is of the form
$\sigma(a,b,c,d)(1,a,b)\otimes(1,c,d)$, with only a subset, ${\cal A}$, of $a,d,c,d$ allowed, and $\sigma(a,b,c,d)=\pm1$. If, say,  $v_x$ 
belongs to $H_o$,
then   the sum of its expansion coefficients, $t_{abcd}$ , with respect to the basis $B_o$, given by $(1,a,b)\otimes(1,c,d)$, with $a,b,c,d$ in $\cal A$,
$\sum_{(a,b,c,d)\in A}{t_{abcd}}$, is equal to one, and a similar sum for $-v_x$ must give $-1$. However, the coefficients with respect to $B$, denoted as $t'_{abcs}$, 
are therefore given by $t'_{abcd}=\sigma(a,b,c,d)t_{abcd}$. Therefore, by Fact 3, $t'_{abcd}-t_{abcd}=0,\pm2,\pm4,...$, and thus
the value of 
$\sum_{(a,b,c,d)\in A}{t'_{abcd}}$ can differ from $\sum_{(a,b,c,d)\in A}{t_{abcd}}$ by $0, \pm2, \pm4,...$. 
I.e. either $\sum_{(a,b,c,d)\in A}{t'_{abcd}}=\pm1$, which is impossible by our assumption that both $\pm v_x$ do not 
belong to $H$, or $|\sum_{(a,b,c,d)\in A}{t'_{abcd}}|\geq 3$. This implies that the value of (\ref{OUTSIDE}), which is non-zero 
for both vertices $v_x$ and $-v_x$, is negative for one of them and positive for the other. That is, they  are in the  interiors of the two different halfspaces separated by $H$. 
I.e., H cannot contain  a face of the polytope. It cuts through it. QED

The conclusion of this section is the following one. 
{\em All faces of the polytope contain 16 vertices.} 

\subsection{Bell inequalities}
A straightforward application of the formula (\ref{DETERMINANT}), and of the other above results implies that
 the tight Bell inequalities are obtainable by the following procedure\footnote{A Bell inequality is tight if it is saturated by all vertices of a face .}.
If we find  nine vertices, which generate a hyperplane containing a given face of the polytope, and arrange them is a sequence in such a way that
 $d[v_1,v_2,....,v_9]>0$, then the (tight) Bell inequality reads
\begin{equation}
 D[E,v_1,v_2,...,v_9]\geq0,
 \end{equation}
where $E=(E_{00},E_{01}, E_{02}, E_{10}, E_{11}, E_{12}, E_{20},E_{21}, E_{22})$. 
The tensor products defining the vertices, $v=(x,y,z)\otimes(w,v,t)$,
are written in a convention consistent with the one for $E$, that is 
$v=(xw,xv,xt,yw,yv,yt,zw,zv,zt)$. 

\section{Bell inequalities: the explicit form}
In order to show how explicitly such inequalities look like, and to have an easy 
method to pinpoint them, one should still:
\begin{itemize}
\item
Show how to find which of the sets of 16 vertices belong to one face.
\end{itemize}

\subsection{Basis sets of nine}
If we have 16 vertices of one face, then a basis generating a hyperplane containing all these vertices is their  arbitrary  subset of nine linearly independent vertices. Therefore we can always choose a product basis of the form $\sigma(a,b,c,d)(1,a,c)\otimes(1,c,d)$, with $(a,b)\neq(1,1)$ and $(c,d)\neq(1,1)$, where $\sigma(a,b,c,d)=\pm1$.
The only question left is, for which sign functions, $\sigma(a,b,c,d)$, such a set of nine generates a hyperplane containing a face.

\subsection{How to find the sets of 16 vertices that generate Bell inequalities (polytope faces)}

Definitely the set  $(1,a,b)\otimes(1,c,d)$, with $a,b,c,d=\pm1$, is in one face of the polytope. Any other set of 16 vertices that 
does not contain oppositely pointing pairs of vertices,
has vertices of the following form $\sigma(a,b,c,d)(1,a,b)\otimes(1,c,d)$, with $a,b,c,d=\pm1$, and a specific ``sign"' function $\sigma(a,b,c,d)=\pm1$. The problem is to decide whether the given set belongs to a single face of the polytope.

We shall show that {\em if and only if $\sigma(a,b,c,d)=S(a,b,c,d)$ where  $S(a,b,c,d)$ is such that $\sum_{a,b}ab S(a,b,c,d)=0$ and $\sum_{c,d}cd S(a,b,c,d)=0$, then the full set of 16 vertices
 is in one hyperplane (which in turn means that they belong to one face)}.
 {Of course the two conditions warrant also that one has, e.g., $\sum_{a,b,c}abc S(a,b,c,d)=0$ and $\sum_{a,b,c,d}abcd S(a,b,c,d)=0$. Therefore $S(a,b,c,d)$
 must have the following form: 
 \begin{equation}
 S(a,b,c,d)=X+Aa+Bb+Cc+Dd+Eac+Fad+Gbc+Hbd, \label{GOOD-SIGNFUNCTIONS}
 \end{equation}
 where $X,A,....,H$ are constants. }
 
 Proof: For a basis we can choose a specific basis set,
 that is $S(a',b',c',d')(1,a',b')\otimes(1,c',d')$, which excludes the seven verices $S(1,1,c,d)(1,1,1)\otimes(1,c,d)$, and 
 $S(a,b,1,1)(1,a,b)\otimes(1,1,1)$.
 
 Let us start with the trivial case of $S(a,b,c,d)=1$. Linear combinations of the the selected nine which give the other seven vertices, 
 have always the property that the sum of the expansion coefficients is $1$. E.g.,  $-\sum_{(a',b')\neq(1,1)}a'b'(1,a',b')=(1,1,1)$, and therefore
 the sum of the coefficients is   $-\sum_{(a',b')\neq(1,1)}a'b'=1$, and therefore $(1,1,1)\otimes(1,c',d')$ has the required property. Similarly \begin{eqnarray}\sum_{(a',b')\neq(1,1),(c',d')\neq(1,1)}a'b'c'd'(1,a',b')\otimes(1,c',d')=(1,1,1)\otimes(1,1,1),\end{eqnarray}
  and $\sum_{(a',b')\neq(1,1),(c',d')\neq(1,1)}a'b'c'd'=1$.
 
 Now we shall relate the above specific case with the general one. Since trivially 
 \begin{eqnarray}&\sum_{(a',b')\neq(1,1),(c',d')\neq(1,1)}a'b'c'd'S(a',b',c',d')[S(a',b',c',d')(1,a',b')\otimes(1,c',d')]&
 \nonumber \\
 &=S(1,1,1,1)S(1,1,1,1)(1,1,1)\otimes(1,1,1),&\end{eqnarray}
 therefore the sum of the expansion coefficients for the vertex $S(1,1,1,1)(1,1,1)\otimes(1,1,1)$ is given by 
 $\sum_{(a',b')\neq(1,1),(c',d')\neq(1,1)}a'b'c'd'S(a',b',c',d')S(1,1,1,1)$. But since  $\sum_{a,b}ab S(a,b,c,d)=0$ and $\sum_{c,d}cd S(a,b,c,d)=0$, one has 
 \begin{equation}
 -\sum_{(a',b')\neq (1,1)}a'b' S(a',b',c,d)=S(1,1,c,d),
 \end{equation}
 and
 \begin{equation}
 -\sum_{(c',d')\neq (1,1)}c'd' S(a,b,c',d')=S(a,b,1,1).
 \end{equation}
 Therefore
 \begin{eqnarray}
 &\sum_{(a',b')\neq(1,1),(c',d')\neq(1,1)}a'b'c'd'S(a',b',c',d')S(1,1,1,1)&
\nonumber \\
 & =-S(1,1,1,1)\sum_{(c',d')\neq(1,1)}S(1,1,c',d')c'd'=[-S(1,1,1,1)]^2=1.   &
\nonumber\\ 
 \end{eqnarray}  
 Similarly, one has 
 \begin{eqnarray}
 &\sum_{(c',d')\neq(1,1)}c'd'S(a',b',c',d')[S(a',b',c',d')(1,a',b')\otimes(1,c',d')]&\nonumber \\
 &= -S(a',b',1,1)[S(a',b',1,1)(1,a',b')\otimes(1,1,1)].   &
\nonumber\\ 
\end{eqnarray}
 The sum of the expansion coefficients for the vertex $S(a',b',1,1)(1,a',b')\otimes(1,1,1)$
is given by  
  \begin{eqnarray}
 &-\sum_{(c',d')\neq(1,1)}c'd'S(a',b',c',d')S(a',b',1,1)= S(a',b',1,1)^2=1,  &
\nonumber\\ 
\end{eqnarray}
i.e. again we have the required property.

In the next step one has to show that if $\sigma(a,b,c,d)$ is not of the form $S(a,b,c,d)$ then the set of sixteen vertices 
does not belong to one face. For the assumed $\sigma$ one must have either $\sum_{a,b}ab\sigma(a,b,c,d)\neq 0$ for some $c,d$, or 
$\sum_{c,d}cd\sigma(a,b,c,d)\neq 0$ for some $a,b$. Assume that the second case holds, for some $a',b'$. Then 
\begin{equation}
-\sum_{(c',d')\neq(1,1)}c'd'\sigma(a',b',c',d')\neq\sigma(a',b',1,1) 
\end{equation}
and thus
\begin{equation}
\sum_{(c',d')\neq(1,1)}c'd'\sigma(a',b',c',d')=x\sigma(a',b',1,1),  
\end{equation}
with $x=1$ or $\pm 3$.
Therefore, since one has
 \begin{eqnarray}
 &\sum_{(c',d')\neq(1,1)}c'd'\sigma(a',b',c',d')[\sigma(a',b',c',d')(1,a',b')\otimes(1,c',d')]&\nonumber \\
 & =-\sigma(a',b',1,1)[\sigma(a',b',1,1)(1,a',b')\otimes(1,1,1)],   &
\nonumber\\ 
\end{eqnarray}
 the sum of the expansion coefficients for the vertex $\sigma(a',b',1,1)(1,a',b')\otimes(1,1,1)$
is given by  
  \begin{eqnarray}
 &-\sum_{(c',d')\neq(1,1)}c'd'\sigma(a',b',c',d')\sigma(a',b',1,1)= -x\sigma(a',b',1,1)^2=-x\neq1,  &
\nonumber\\ 
\end{eqnarray}
i.e. the vertex does not belong to the plane defined by the chosen subset of nine!

\section{Bell inequalities: equivalent form}
Above it was shown what is the full set of Bell inequalities for the considered problem. However, one must admit that the 
the derived form of the inequalities is not good for practical applications.
Below it will be shown that one can write down the inequalities in a much simpler form.

{\em The full set of Bell inequalities, 
forming the necessary condition for the positivity of the unphysical 
distribution, can be
written as} 
\begin{eqnarray}
&2^4-
\sum_{a,b,c,d=\pm1}S(a,b,c,d)\big(E_{00}+E_{01}c+E_{02}d+E_{10}a& \nonumber \\
&+E_{11}ac+
E_{12}ad+E_{20}b+E_{21}bc+E_{22}bd\big)\geq 0.&\nonumber \\
\label{POSITIVITY-SUM-1}
\end{eqnarray}
This can also be written down as:
\begin{eqnarray}
&2^4-
\sum_{a,b,c,d=\pm1}S(a,b,c,d)v_{abcd}\cdot\hat{E}\geq 0&
\label{POSITIVITY-SUM-2}
\end{eqnarray}
where we use the following convention $$v_{abcd}=(1,a,b)\otimes(1,c,d)=(1,c,d,a,ac,ad,b,bc,bd),$$ 
$$\hat{E}=(E_{00},E_{01},E_{02},E_{10},
E_{11},
E_{12},E_{20},E_{21},E_{22}),$$ and $\cdot$ denotes scalar product.

We shall show that for sixteen vertices of the form $S(a,b,c,d)(1,a,b)\otimes(1,c,d)$, i.e., ones defining
a face, one has saturation of the inequality. This proves the equivalence of the inequality (\ref{POSITIVITY-SUM-1})
with the ones discussed in the previous chapters.
Take a specific vertex belonging to the face defined by the sign function $S(a,b,c,d)=X+Aa+Bb+Cc+Dd+Eac+Fad+Gbc+Hbd$, namely 
$$v_{a^ob^oc^od^o}=S(a^o,b^o,c^o,d^o)(1,a^o,b^o)\otimes(1,c^o,d^o)$$
One has 
$$
\sum_{a,b,c,d=\pm1}S(a,b,c,d)v_{abcd}\cdot v_{a^ob^oc^od^o}$$
$$
=S(a^o,b^o,c^o,d^o)\sum_{a,b,c,d=\pm1}(X+Aa+Bb+Cc+Dd+Eac+Fad+Gbc+Hbd)$$
$$
\times (1+a^oa+b^ob+c^oc+d^od+a^oc^oac+a^od^oad+b^oc^obc+b^od^obd)\nonumber \\
=2^4S(a^o,b^o,c^o,d^o)$$
$$
\times(X+Aa^o+Bb^o+Cc^o+Dd^o+Ea^oc^o+Fa^od^o+Gb^oc^o+Hb^od^o)$$
\begin{eqnarray}
&=2^4S(a^o,b^o,c^o,d^o)^2=2^4.&
\label{POSITIVITY-SUM-3}
\end{eqnarray}
That is, the inequality is saturated for all vertices belonging to the face.
Since for vertices that do not belong to the face, that is for $-v_{a^ob^oc^od^o}$, one gets $-2^4$, therefore the inequality is 
a tight Bell inequality associated with the face generated by the sign function $S(a,b,c,d)$.
QED.

\section{Ramifications}
The procedure 
can be generalized to an arbitrary Bell problem involving
$N$ parties, two valued observables, and $3$ local settings per observer. Generalization 
to more than two parties is trivial, and will be presented below.
So is generalization to more than three settings of the derivation of the inequalities as necessary conditions for local realism (proof of their sufficiency may be not easy anymore). 

\subsection{Three or more observers}
For $N>2$ it is straightforward 
to write down the full set of tight Bell inequalities in the form of a single ``synthetic" formula.
E.g., for three parties, dichotomic observables, and three 
settings at each side such a set of tight Bell inequalities is given by
\begin{equation}
-1\leq\sum_{a,b,c,d,e,f=\pm1}q^{[3]}(a,b,c,d,e,f,S^{[3]})\leq1,
\end{equation} 
where 
\begin{itemize}
\item
$S^{[3]}$ is any admissible ``sign" function of the variables $a,b,c,d,e,f$ 
with the property that in its Fourier expansion products $ab$, $cd$ and $ef$
never appear (i.e., all products of indices of a local observer are missing),
\item the following overcomplete bases are selected from the set of 
deterministic 
``orders" sent by the local realistic source to the observers:
\begin{equation}
S^{[3]}(a,b,c,d,e,f)(1,a,b)\otimes(1,c,d)\otimes(1,e,f)=
V^{[3]}_{abcdef,S^{[3]}},
\label{VERTICES-3}\end{equation}
where most importantly the sign function belongs to the class of admissible ones,
\item
and finally 
\begin{equation}
q^{[3]}(a,b,c,d,e,f,S^{[3]})=\hat{E}\cdot V^{[3]}_{abcdef,S^{[3]}},
\label{BELL-GHZ}
\end{equation} 
where the set of three party correlation functions 
$E_{000}, E_{001},...E_{222}$ is written down as a vector (or rather a  three index tensor, $\hat{E}$) in 
$\bf{R}^{27}=\bf{R}^3\otimes\bf{R}^3\otimes\bf{R}^3$.
\end{itemize}
Note that the set of inequalities (\ref{BELL-GHZ}) has never 
appeared in the literature, just like zillions of other ones derivable by this method.

\subsection{Other generalizations}

If one wants to build Bell inequalities involving the full set of 
observable data for the experiment, that is
in the two parties case, also local averages of the local results, 
not only the 
correlation functions of products
results of the two parties, one can re-interpret the inequalities in the following way.

Let us consider the two-observer, {\em two-settings} problem, with dichotomic observables.
Consider the following subset of the vertices of the $3\times3$
problem
\begin{equation}
W'(m_1^{(1)}m_1^{(2)};m_2^{(1)}m_2^{(2)})=(1,m_1^{(1)},m_1^{(2)})
\otimes(1,m_2^{(1)},m_2^{(2)}).
\label{NEWTENSOR}
\end{equation}
Simply the hidden results of the local experiments for the first settings, $m_j^{0}$ with $j=1,2$, was replaced by $1$. 
Now the averages of components of the (\ref{NEWTENSOR}) of such vercites (or if you like convex combinations of all of them), that is
\begin{equation}
\sum_{ W'}P(W')W',
\label{MODEL-W}
\end{equation}
give: normalization condition,  averages of local results, and finally all correlation functions, for all combinations of the two pairs of settings. I.e. nothing is missing 
from the full description of the phenomena, for the problem in question. Thus we get the full description of observable phenomena for an experiment
for which a local realistic model exist. Since the new vertices satisfy the inequalities derived for the $3\times3$ problem 
involving correlation functions, we get immediately a necessary condition for the existence of (\ref{MODEL-W}).  In other words, we get the CH inequalities. When this approach is (straightforwardly) generalized to more than two parties new families of tight Bell inequalities can be derived.

\subsection{Relation with the complete set of two settings inequalities}

The complete set of two setting Bell inequalities for $N$
parties, dichotomic observables and two possible settings for each party, can be recovered from the family of inequalities presented here.
Consider first the two party problem. One can  choose the subset of admissible sign functions of the following form $S(a,b,c,d)=S(a,c)$, i.e. depending on only one index per observer. Due to this fact all components of $S$ in the expansion (\ref{GOOD-SIGNFUNCTIONS}) which contain $b$ and $d$ drop out, and the inequality reduces to a two-setting one.
Similarly, for three parties problem one can choose
$S^{[3]}(a,b,c,d,e,f)=S^{[3]}(a,c,e),$
 with equivalent result, etc. 

\section{Acknowledgments}
Many thanks to prof. N. David Mermin for extended discussions, and especially for demolishing 
the earlier version of the proof. The problem was formulated during author's stay at Erwin Schroedinger Institute for Mathematical Physics (ESI), Vienna.

The work is part of the EU 6FP programmes QAP and  SCALA.  
The earlier stage was supported by the MNiI grant 1 P03B 04927.
Authors acknowledges the Professorial Subsidy of FNP.
Currently the author is supported by Wenner Gren Foundation.

\end{document}